\documentclass{PoS}
\usepackage{slashed}

\newcommand{\sea}{\mathrm{sea}}

\title{Vector current renormalisation in momentum subtraction schemes using the HISQ action}

\ShortTitle{Vector current renormalisation in momentum subtraction schemes using the HISQ action}

\author{\speaker{D. Hatton}\\
        SUPA, School of Physics and Astronomy, University of Glasgow, Glasgow, G12 8QQ, UK\\
        E-mail: \email{d.hatton.1@research.gla.ac.uk}}

\author{C. T. H. Davies\\
        SUPA, School of Physics and Astronomy, University of Glasgow, Glasgow, G12 8QQ, UK\\
        E-mail: \email{christine.davies@glasgow.ac.uk}}

\author{G. P. Lepage\\
        Laboratory for Elementary-Particle Physics, Cornell University, Ithaca, New York 14853, USA\\
        E-mail: \email{g.p.lepage@cornell.edu}}

\author{A. T. Lytle\\
        INFN, Sezione di Roma Tor Vergata, Via della Ricerca Scientifica 1, 00133 Roma RM, Italy\\
        E-mail: \email{andrew.lytle@roma2.infn.it}}

\author{HPQCD Collaboration\\
        www.physics.gla.ac.uk/HPQCD}

\abstract{As the only lattice vector current that does not require renormalisation is the point-split conserved current it is convenient to have a robust, precise and computationally cheap methodology for the calculation of vector current renormalisation factors, $Z_V$. Momentum subtraction schemes, such as RI-SMOM, implemented nonperturbatively on the lattice provide such a method if it can be shown that the systematic errors, e.g. from condensates, are well controlled. 

We present $Z_V$ calculations for the conserved current in both the RI-SMOM and RI$'$-MOM momentum subtraction schemes as well as local current renormalisation in the RI-SMOM scheme. By performing these calculations at various values of the momentum scale $\mu$ and different lattice spacings we can investigate the presence of power suppressed nonperturbative contributions and compare the results to expectations arising from the Ward-Takahashi identity. Our results show that the RI-SMOM scheme provides a well controlled determination of $Z_V$ but the standard RI$'$-MOM scheme does not.

We then present some preliminary uses of these $Z_V$ calculations in charm physics.}

\FullConference{37th International Symposium on Lattice Field Theory - Lattice2019\\
    16-22 June 2019\\
    Wuhan, China}

\begin{document}

\section{Motivation}

While there exists a conserved vector current on the lattice that does not require 
renormalisation it can be complicated to implement for highly improved actions such 
as HISQ \cite{Follana:2006rc} where it includes both 1-link and 3-link pieces. It is therefore more 
practically convenient to work with the simple local vector current which then requires 
a renormalisation calculation to be performed. The HPQCD collaboration has previously 
used the calculation of form factors at $q^2 = 0$ to renormalise the local vector 
current across a range of lattice spacings ($\sim 0.15 - 0.09$ fm) \cite{Chakraborty:2017hry}. These results achieved high precision but to do so 
required the computation of two and three-point functions on large numbers of configurations 
($\mathcal{O}(1000)$) at each lattice spacing. For future work that will require further 
renormalisation calculations to be performed (for example, on finer lattices) it is desirable 
to have a numerically faster method. Such a method is offered by the use of momentum 
subtraction schemes implemented on the lattice with momentum sources. It is then found that 
good precision can be obtained with only $\sim20$ configurations.

Here we perform an analysis of the systematic effects, in particular nonperturbative (condensate) 
contributions, present in momentum subtraction scheme calculations of vector current 
renormalisation.

\section{The Ward-Takahashi identity}

On the lattice there exists an exact Ward-Takahashi identity (WTI) which relates the
finite difference operator acting on a matrix element involving the (action dependent) conserved vector
current to a difference of propagators. We have

\begin{equation} \label{eq:lat-ward-pos}
  \langle \Delta_{\mu} J_{\mathrm{con}}^{\mu}(x) \psi(y_1)\overline{\psi}(y_2) \rangle = \delta_{y_2,x} 
  \langle \psi(y_1)\overline{\psi}(x) \rangle - \delta_{y_1,x} \langle \psi(x)
  \overline{\psi}(y_2) \rangle .
\end{equation}

Once the discrete Fourier transform of Eq.~\ref{eq:lat-ward-pos} has been taken the lattice WTI
becomes ($\tilde{x}$ is the mid-point of $x$ and $x+\hat{\mu}$)

\begin{equation} \label{eq:lat-ward-mom}
  (1-e^{iaq_{\mu}}) \sum_x \langle \psi(p_1) e^{-iq \cdot x} J^{\mu}_{\mathrm{con}}(\tilde{x}) \overline{\psi}(p_2) 
  \rangle = \langle \psi(p_1) \overline{\psi}(p_2) \rangle - \langle \psi(p_2) \overline{\psi}(p_1) \rangle .
\end{equation}

In the continuum this identity means that once wavefunction renormalisation has been performed it is not 
possible to separately renormalise the vector current and therefore $Z_V = 1$. As Eq.~\ref{eq:lat-ward-mom} 
is exact as a fully nonperturbative expression on the lattice, this is also true of the lattice conserved vector current. 
It also means that methods for determining $Z_V$ for non-conserved currents that make use of the Ward-Takahashi identity are 
protected against contamination by nonperturbative artefacts. This is the case with the RI-SMOM scheme.

\section{RI-SMOM and RI$'$-MOM}

The RI-SMOM (shortened here to SMOM) scheme is defined using quark propagators and 
vertex functions of operators between external off-shell quark states. Spin-colour traces over 
projected pieces of the inverse propagator and vertex functions are compared to their tree level values to define their 
renormalisation constants. The SMOM kinematic setup inserts momentum at the vertex where the 
operator is placed so that the incoming and outgoing quark momenta are not equal $q=p_1 - p_2 
\neq 0$. These momenta are chosen to be in a symmetric configuration with $p_1^2 = p_2^2 = q^2 
\equiv \mu^2$ \cite{Sturm:2009kb}.

The vector vertex function is constructed from quark 
fields $\psi$ as:

\begin{equation}
G_V = \langle \psi(p_1) (\sum_x \overline{\psi}(x) \gamma_{\mu} \psi(x) e^{i(p_1-p_2)\cdot x} ) \overline{\psi}(p_2) \rangle . \nonumber
\end{equation}

After amputation by the quark propagator $S$,  $\Lambda_V = S^{-1}(p_1) G_V S^{-1}(p_2)$, 
this vertex function is used to define the vector current renormalisation.  
$Z_V=Z_q/\mathrm{Tr}(P_V\Lambda_V)$ with $Z_q = \mathrm{Tr}(\slashed{p}S^{-1}(p))$ and $P_V = (1/12)q_{\mu}\slashed{q}$ in the SMOM scheme (in the continuum). 
Here the trace is over spin and colour indices.

If we consider a lattice form of the SMOM $Z_V$ definition where $\Lambda_V$ is now the amputated 
vector vertex function for the lattice conserved vector current we can amputate the WTI and multiply 
by $(1/12\hat{q}^2)\slashed{\hat{q}}$ to obtain (ignoring subtleties related to our use of staggered quarks \cite{Lytle:2013qoa,Lytle:2018evc})

\begin{equation} \label{eq:SMOM-condition}
	\frac{Z_q}{Z_V} = \frac{1}{12\hat{q}^2}\mathrm{Tr} \left(\frac{-2i}{a}\mathrm{sin}(aq_{\mu}/2)\Lambda_V^{\mu} \slashed{\hat{q}} \right)
	= \frac{1}{12\hat{q}^2} [\mathrm{Tr} (S^{-1}(p_2) \slashed{\hat{q}}) - \mathrm{Tr}(S^{-1}(p_1)\slashed{\hat{q}})] .
\end{equation}

The $\hat{q}$ in $\slashed{\hat{q}}$ is a discretisation of $q$ chosen so that $Z_q$ is 1 in the free theory, $\hat{q}=\mathrm{sin}(aq)+(1/6)\mathrm{sin}^3(aq)$ \cite{Lytle:2013qoa}. 

The r.h.s of Eq.~\ref{eq:SMOM-condition} is equal to $(1/12q^2)\mathrm{Tr}(S^{-1}(\hat{q})\slashed{\hat{q}}) 
= Z_q$ in the continuum. If this remains unbroken by discretisation effects on the lattice then 
$Z_V=1$ for the conserved current in the SMOM scheme independent of mass, momentum and lattice spacing. Through explicit numerical 
calculation we see this to be true up to the 0.05\% level of our statistical errors.

The RI$'$-MOM scheme \cite{Chetyrkin:1999pq} uses a kinematic setup in which no momentum is inserted 
at the vertex ($q=0$) and there is therefore only one momentum in the problem, $p$. The definition of $Z_V$ 
is different in RI$'$-MOM although the same $Z_q$ definition is used. The difference is that 
projector $P_V$ in RI$'$-MOM is just $\gamma_{\mu}$ which makes its application a little simpler. 
However, as the WTI is \textit{not} used in the construction of the scheme to protect $Z_V$ for the conserved current against 
renormalisation $Z_V$ is not equal to 1 for the conserved current, even in the continuum, and a perturbative matching to the 
$\overline{\mathrm{MS}}$ scheme is required \cite{Chetyrkin:1999pq}. We may also expect nonperturbative 
effects to be present in $Z_V$ calculated on the lattice using the RI$'$-MOM scheme even for the conserved current as the WTI does not guarantee their cancellation, as seen in 
Figure 1.

\section{Results}

The details of our implementation can largely be found in \cite{Lytle:2018evc} and \cite{Lytle:2013qoa}. We stringently fix to Landau 
gauge and calculate propagators from momentum 
sources $e^{-ip\cdot x}$ on 20 configurations. The results presented here use 
the MILC 2+1+1 ensembles listed in Table 1 \cite{Bazavov:2012xda} where the $Z_V$ numbers from set 4 are 
preliminary.

\begin{table}
\label{tab:ensembles}
\caption{MILC 2+1+1 ensembles used in the analysis presented here. The results on set 4 are preliminary.}
\centering
\begin{tabular}{llllllll} 
\hline \hline
Set & $\beta$ & $a$ [fm] & $L_s$ & $L_t$ & $am_l^{\sea}$ & $am_s^{\sea}$ & $am_c^{\sea}$ \\
\hline
1 & 6.0 & 0.12404(66) & 24 & 64 & 0.0102 & 0.0509 & 0.635 \\
2 & 6.30 & 0.08872(47) & 48 & 96 & 0.00363 & 0.0363 & 0.430 \\
3 & 6.72 & 0.05922(33) & 48 & 144 & 0.0048 & 0.024 & 0.286 \\
4 & 7.00 & 0.04406(23) & 64 & 192 & 0.00316 & 0.0158 & 0.188 \\
\hline \hline
\end{tabular}
\end{table}

\subsection{RI$'$-MOM: $Z_V$ for the conserved current}

We calculate the vector current renormalisation in the RI$'$-MOM scheme for three different lattice 
spacings and four different $\mu$ values, apply the matching factor factor to $\overline{\mathrm{MS}}$ \cite{Chetyrkin:1999pq,Gracey:2003yr} 
known through order $\alpha_s^3$ and fit the resulting data. We use the fit function 

\begin{eqnarray} \label{eq:mom-con-fit}
	Z_V^{\mathrm{con}}(a,\mu) = 1 + \sum_i c_{a^2\mu^2}^{(i)}(a\mu / \pi)^{2i} + \sum_i c_{\alpha 
	a^2\mu^2}^{(i)}(a\mu / \pi)^{2i} \alpha_{\overline{\mathrm{MS}}}(1/a) + \\ \sum_j c_{\mathrm{cond}}^{(j)} 
	\alpha_{\overline{\mathrm{MS}}}(\mu) \frac{(1\ \mathrm{GeV})^{2j}}{\mu^{2j}} \times [1+
	c_{\mathrm{cond},a^2}^{(j)}(a\Lambda/\pi)^2 ] + c_{\alpha} \alpha_{\overline{\mathrm{MS}}}^4(\mu) 
	\nonumber.
\end{eqnarray}

This includes discretisation errors, power suppressed condensate terms (expected to arise here as already discussed) and a term to account for the uncertainty in the 
$\overline{\mathrm{MS}}$ matching at order $\alpha_s^4$. Here we are using the conserved current so $\alpha_s(1/a)$ terms not multiplied by discretisation terms should not appear.
The RI$'$-MOM conserved $Z_V$ results and the fit using Eq.~\ref{eq:mom-con-fit} ($\chi^2/\mathrm{d.o.f} = 0.6$) are shown in Figure 1.

\begin{figure} \label{fig:mom-con}
	\centering
	\includegraphics[width=0.6\textwidth]{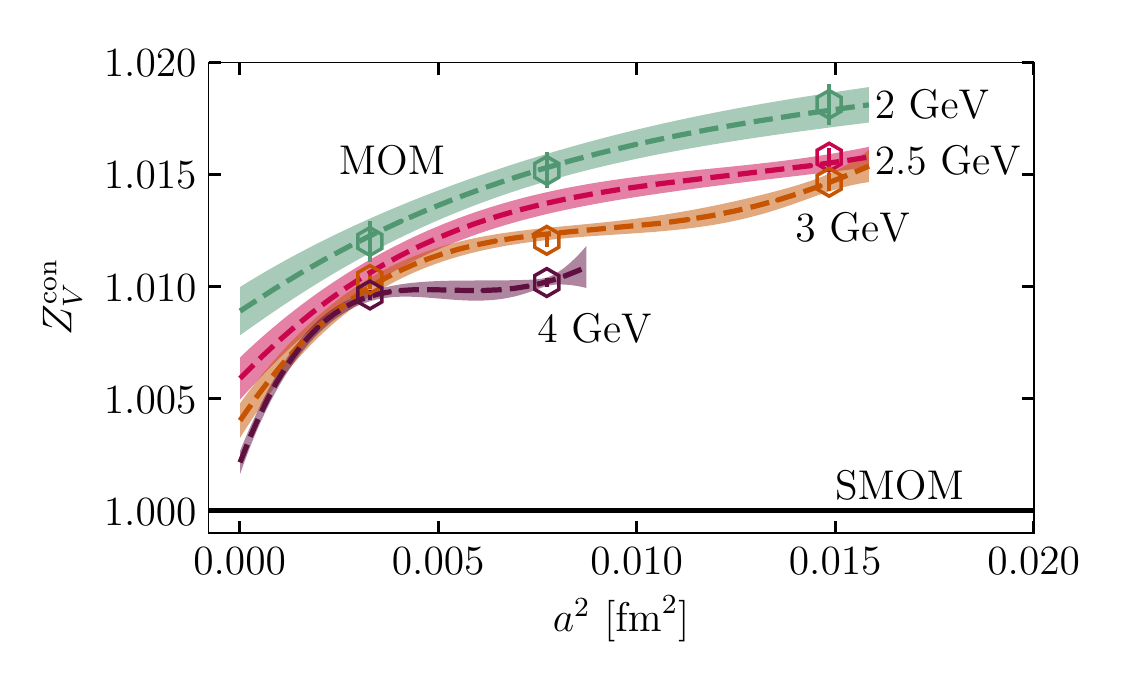}
	\caption{The renormalisation constant of the conserved vector current in the RI$'$-MOM scheme for 
	different values of the momenta $\mu$ and different lattice spacings. There is clearly dependence 
	on both quantities. In addition to this there are signs of a $\sim$1\% condensate effect demonstrated by 
	the disagreement between the continuum extrapolations for different $\mu$. This is to be compared 
	to the RI-SMOM scheme where the value of the conserved $Z_V$ is 1 (the black line), independent of $\mu$ or $a^2$ as observed through explicit computation.}
\end{figure}

A good $\chi^2$ cannot be obtained if the condensate terms are omitted. The presence of these terms in 
the $Z_V$ determined for the conserved current means that the continuum limit depends (incorrectly) on 
$\mu$. For example, at $\mu=2$ GeV the condensate terms correspond to a $\sim$1\% systematic error in 
$Z_V$. Such effects are absent in RI-SMOM, shown by the straight line at value 1 in Figure 1.

\subsection{RI-SMOM: $Z_V$ for the local current}

The local vector current is not conserved but, once renormalised, should have the same matrix elements as 
the conserved current in the continuum limit. The fact that $Z_V=1$ for the conserved current in the 
RI-SMOM scheme implies that the renormalisation factor for the local current must be free of condensate 
contributions in the continuum limit. We can test this by studying the behaviour of the SMOM $Z_V$ for the local current that we calculate. HPQCD has previously calculated the local current 
renormalisation using form factor methods \cite{Chakraborty:2017hry}. The difference between the results for $Z_V$ 
from the form factor method (denoted $Z_V^{F(0)}$) and the SMOM scheme should be purely a consequence of 
discretisation effects; the perturbative QCD series that renormalises the local to the conserved current should 
cancel between the two and neither should have condensate contributions (this was tested for the form factor method in 
\cite{Chakraborty:2017hry}). Our results for the difference of $Z_V$ values are shown in Figure 2 along with 
a fit to a polynomial of even powers of $a\mu$ (we also include powers of $a\mu$ multiplied by $\alpha_s$). The fit works 
well, demonstrating the absence of condensate effects for the local vector current renormalisation in the RI-SMOM scheme.

\begin{figure} \label{fig:SMOM-FF}
  \centering
  \includegraphics[width=0.6\textwidth]{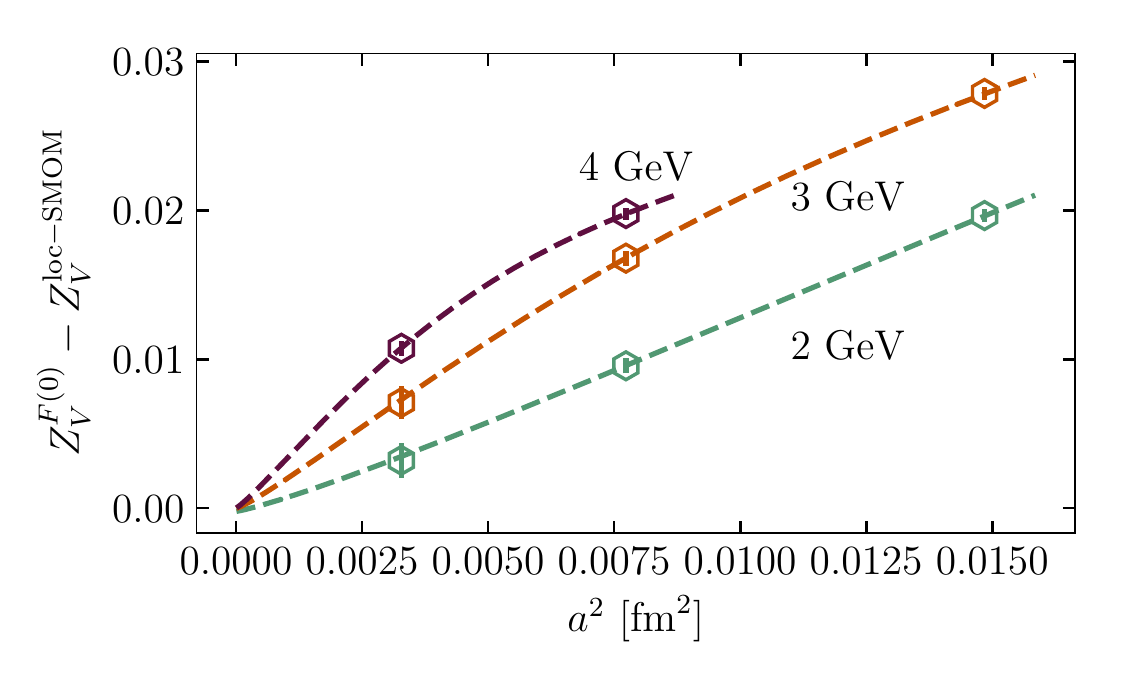}
	\caption{The difference of the local current renormalisation factors determined from vector form
  factors and in the RI-SMOM scheme. The fit includes just discretisation effects, indicating the absence
  of condensate effects in the vector current renormalisation calculated in the RI-SMOM scheme as well as 
	the consistency between the two sets of results.}
\end{figure}

\section{Applications to charm physics}

These local vector current renormalisations, calculated in the RI-SMOM scheme, may be used in conjunction 
with existing HPQCD charmonium data on the MILC 2+1+1 ensembles \cite{Galloway:2014tta}. The vector renormalisation is required 
for the computation of the $J/\psi$ decay constant. A preliminary continuum extrapolation of the renormalised
$J/\psi$ decay constant is shown in Figure 3 where good agreement with the PDG \cite{Tanabashi:2018oca} value can be 
seen, determined from the $J/\psi$ leptonic width. This imporves on HPQCD's earlier results on $n_f=$2+1 gluon configurations \cite{Donald:2012ga}.

\begin{figure} \label{fig:fjpsi}
  \centering
  \includegraphics[width=0.6\textwidth]{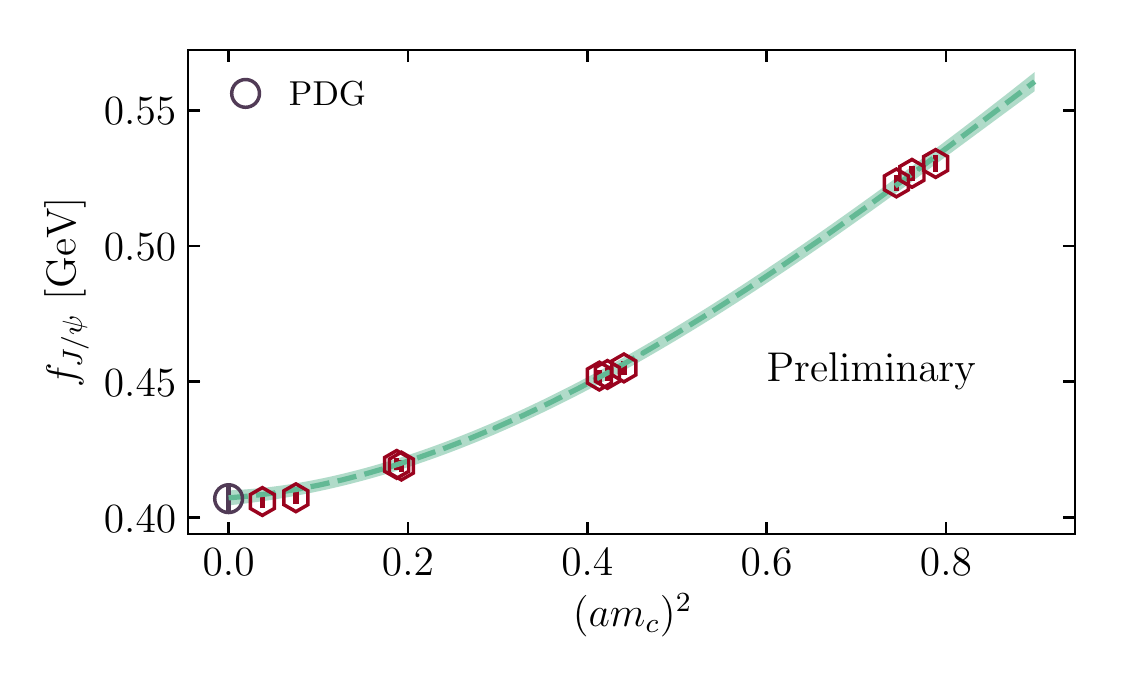}
	\caption{Preliminary continuum extrapolation of the $J/\psi$ decay constant using $Z_V$ factors in the
  RI-SMOM scheme.}
\end{figure}

We may also use the same correlators to calculate vector correlator time moments defined by
$\mathcal{M}_n = Z_V^2 \sum_t t^n C_{J/\psi}(t)$,
with $C_{J/\psi}(t)$ denoting the vector correlator at the charm mass. The continuum extrapolations 
of these moments can be compared to determinations using experimental data \cite{Chetyrkin:2009fv}. This is done on the left-hand side of 
Figure 4. These time moments can be used to calculate the charm connected 
contribution to the leading order hadronic vacuum polarisation contributiuon to the anomalous 
magnetic moment of the muon $a_{\mu}^c$ \cite{Chakraborty:2014mwa}. This can be done at each lattice spacing and extrapolated to 
the continuum or each time moment can be extrapolated to $a=0$ individually and then combined. 
The right-hand side of Figure 4 compares these two methods as well as providing comparison to a previous HPQCD \cite{Chakraborty:2014mwa} and 
a BMW \cite{Borsanyi:2017zdw} determination of the same quantity.

\begin{figure} \label{fig:time-moments}
  \centering
  \includegraphics[width=0.45\textwidth]{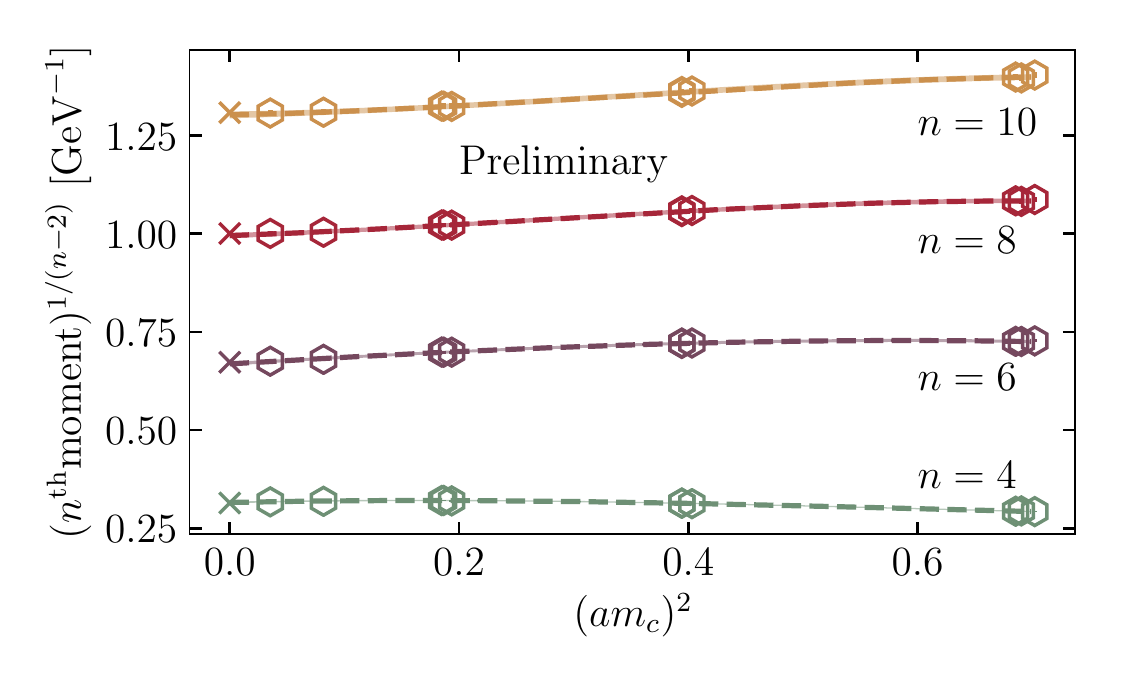}
	\includegraphics[width=0.45\textwidth]{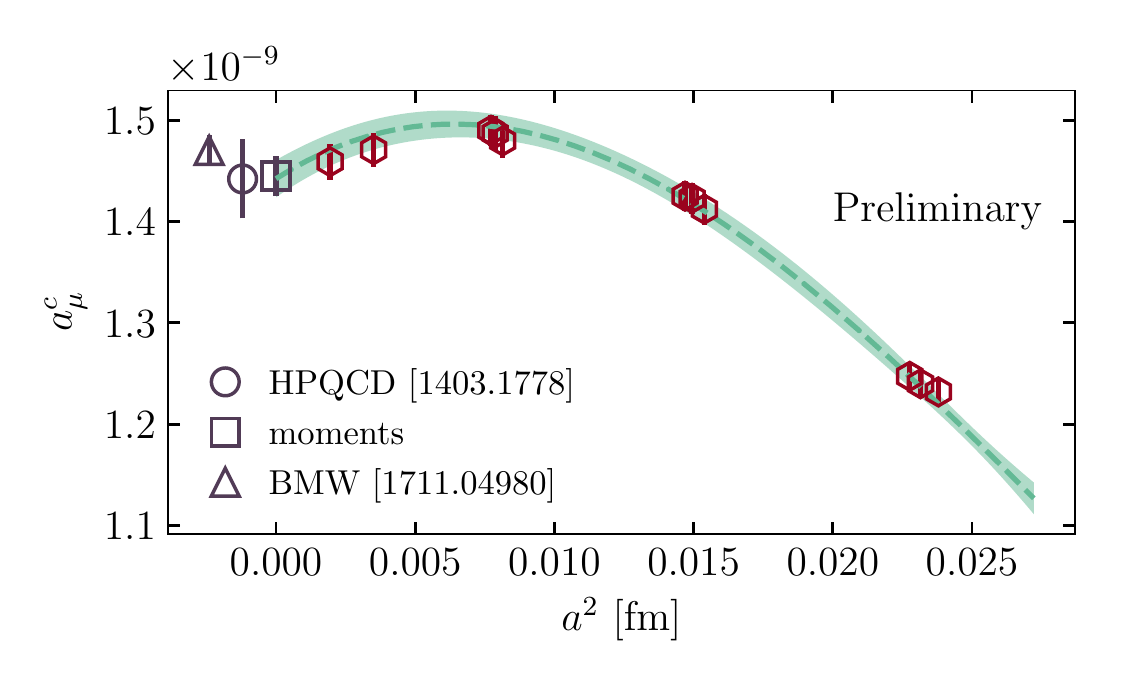}
	\caption{Left: Preliminary continuum extrapolations of various vector time moments compared to their
  extraction from experimental data. Right: Preliminary results for $a_{\mu}^c$ (using $Z_V$ calculated in the SMOM scheme) with 
 comparisons to previous results.}
\end{figure}


\section{Conclusion}

We have performed a numerical implementation of the HISQ conserved current which we have used to 
explicitly show that $Z_V=1$ for the conserved current in RI-SMOM with no visible condensate 
effects. In comparison, there are $\sim1\%$ nonperturbative effects which contaminate results for the 
conserved current in the RI$'$-MOM scheme. This is to be expected from the different construction of the two 
schemes: RI-SMOM makes use of the Ward-Takahashi identity to protect the conserved vector current 
while RI$'$-MOM does not. We have also demonstrated consistency between SMOM and form factor results 
for the local vector current renormalisation indicating a lack of nonperturbative effects in this 
case as well. This implies that vector current renormalisations calculated in the SMOM scheme may be used in 
lattice calculations without needing to worry about nonperturbative contamination. The same cannot be 
said of the RI$'$-MOM scheme. With that in mind we have demonstrated some preliminary applications 
of SMOM local vector current renormalisations to charm physics.

\vspace{0.25cm}

\textbf{Acknowledgments} We are grateful to MILC for the use of their gluon field ensembles. This
work was supported by the UK Science and Technology Facilities Council. The calculations used
the DiRAC Data Analytic system at the University of Cambridge, operated by the University of
Cambridge High Performance Computing Service on behalf of the STFC DiRAC HPC Facility
(www.dirac.ac.uk). This is funded by BIS National e-infrastructure and STFC capital grants and
STFC DiRAC operations grants.

\end{document}